\documentclass[aps,prb,reprint,superscriptaddress]{revtex4-2}
\usepackage{graphicx}
\usepackage{amsmath}
\usepackage{amssymb}
\usepackage{color}
\usepackage{hyperref}
\usepackage{upgreek}

\usepackage{easyReview}

\begin{document}
\title{Design, synthesis, and physical properties of the intergrowth compound Eu$_2$CuZn$_2$As$_3$}
\author{Xiyu Chen}
\altaffiliation{Contributed equally to this work}
\affiliation{Key Laboratory of Quantum Materials and Devices of Ministry of Education, School of Physics, Southeast University, Nanjing 211189, China}
\author{Ziwen Wang}
\altaffiliation{Contributed equally to this work}
\affiliation{Key Laboratory of Quantum Materials and Devices of Ministry of Education, School of Physics, Southeast University, Nanjing 211189, China}
\author{Wuzhang Yang}
\affiliation{School of Science, Westlake University, Hangzhou 310024, China}
\affiliation{Institute of Natural Sciences, Westlake Institute for Advanced Study, Hangzhou 310024, China}
\author{Jia-Yi Lu}
\affiliation{School of Physics, Interdisciplinary Center for Quantum Information and State Key Laboratory of Silicon and Advanced Semiconductor Materials, Zhejiang University, Hangzhou 310058, China}
\author{Zhiyu Zhou}
\affiliation{Key Laboratory of Quantum Materials and Devices of Ministry of Education, School of Physics, Southeast University, Nanjing 211189, China}
\author{Shanshan Wang}
\affiliation{Key Laboratory of Quantum Materials and Devices of Ministry of Education, School of Physics, Southeast University, Nanjing 211189, China}
\author{Zhi Ren}
\affiliation{School of Science, Westlake University, Hangzhou 310024, China}
\affiliation{Institute of Natural Sciences, Westlake Institute for Advanced Study, Hangzhou 310024, China}
\author{Guang-Han Cao}
\affiliation{School of Physics, Interdisciplinary Center for Quantum Information and State Key Laboratory of Silicon and Advanced Semiconductor Materials, Zhejiang University, Hangzhou 310058, China}
\affiliation{Collaborative Innovation Center of Advanced Microstructures, Nanjing University, Nanjing 210093, China}
\author{Shuai Dong}
%\email{sdong@seu.edu.cn}
\affiliation{Key Laboratory of Quantum Materials and Devices of Ministry of Education, School of Physics, Southeast University, Nanjing 211189, China}
\author{Zhi-Cheng Wang}
\email{wzc@seu.edu.cn}
\affiliation{Key Laboratory of Quantum Materials and Devices of Ministry of Education, School of Physics, Southeast University, Nanjing 211189, China}
\date{\today}

\begin{abstract}

The rational combination of existing magnetic topological compounds presents a promising route for designing new topological materials. We report the synthesis and comprehensive characterization of the layered quaternary intergrowth compound Eu$_2$CuZn$_2$As$_3$, which combines structural units of two known magnetic topological materials, EuCuAs and EuZn$_2$As$_2$. Eu$_2$CuZn$_2$As$_3$ exhibits an antiferromagnetic ground state with successive magnetic transitions: quasi-two-dimensional ordering at $T_\mathrm{M} = 29.3$\,K, long-range antiferromagnetic ordering at $T_\mathrm{N} = 19$\,K, and spin-reorientation at $T_\mathrm{SR} = 16.3$\,K. The stepwise magnetic transitions manifest as plateau-like anomalies in the heat capacity. These transitions originate from multiple superexchange pathways and periodic variation of interplane Eu-Eu distances in the intergrowth structure. Charge transport shows a pronounced resistivity increase above $T_\mathrm{N}$ followed by minimal change below the ordering temperature. Magnetic fields rapidly suppress this resistivity rise, yielding significant negative magnetoresistance. Remarkably, Eu$_2$CuZn$_2$As$_3$ inherits the nonlinear anomalous Hall effect characteristic of its parent compounds. Energy evaluations of collinear spin configurations reveal a lowest-energy state with ferromagnetic coupling between Eu planes in EuCuAs units while maintaining antiferromagnetic coupling within EuZn$_2$As$_2$ units. The corresponding electronic structure displays potentially topologically nontrivial features. Our work demonstrates the efficacy of structural hybridization for discovering novel magnetic topological materials and establishes a general strategy for materials discovery.

\end{abstract}
\maketitle

\section{INTRODUCTION}
%创新点：递进逐次的转变

Intrinsic magnetic topological materials leverage the strong coupling between inherent magnetism and topologically nontrivial electronic states to enable the realization of topologically protected edge states and the quantum anomalous Hall effect without requiring external magnetic fields or additional doping~\cite{armitageWeylDiracSemimetals2018,tokuraMagneticTopologicalInsulators2019,bernevigProgressProspectsMagnetic2022,liProgressAntiferromagneticTopological2024}. These remarkable properties exhibit tremendous potential for applications ranging from information storage and topological quantum computing to low-power spintronic devices~\cite{smejkalTopologicalAntiferromagneticSpintronics2018,heTopologicalSpintronicsMagnetoelectronics2022,duanTopologicalQuantumMaterials2024,nayakNonAbelianAnyonsTopological2008,liuMagneticTopologicalInsulator2023}. However, accurately screening intrinsic magnetic topological materials and predicting their properties remains challenging for current theoretical approaches due to the complex interplay between magnetic interactions and other quantum effects~\cite{changBandEngineeringDirac2015,tangEfficientTopologicalMaterials2019,choudharyHighthroughputSearchMagnetic2021,suHighthroughputFirstprinciplePrediction2022,xuDiscoveringTwodimensionalMagnetic2024}. This complexity significantly hinders the efficient discovery of novel magnetic topological materials. A promising strategy to address this challenge involves hybridizing materials that exhibit strongly coupled magnetic order and topological states. For layered compounds specifically, this can be achieved through alternating stacking of structural motifs from parent materials to construct new compounds with intergrowth structures~\cite{wanLayeredHybridSuperlattices2024,baiBulkSuperlatticeAnalogues2022}.

In recent years, Eu-containing Zintl compounds have garnered significant attention owing to the strong interplay between their narrow band gaps and magnetism, which yields highly tunable transport properties, magnetic behavior, and electronic band topology~\cite{5EuCd2P2_AM,9EuSn2As2,8EuIn2As2_topologicalinsulator,7EuIn2As2_Diraccones,10Eu5In2As6,zhouUnusualMagneticTransport2024,rosaColossalMagnetoresistanceNonsymmorphic2020,36EuZn2P2CXY,35EuCd2P2CXY}. Among these, two classes of Eu-based layered compounds, the CaAl$_{2}$Si$_{2}$-type Eu$M_{2}X_{2}$ ($M$ = Cd, Zn; $X$ = P, As, Sb) and SrPtSb-type Eu$TX$ ($T$ = Cu, Ag, Au; $X$ = P, As, Sb, Bi), are recognized as candidate magnetic topological materials~\cite{chenRecentAdvancesUnderstanding2024,1EuCd2As2_SA,4PRL_EuCu2As2,17EuZn2Sb2,suMagneticExchangeInduced2020,24EuZn2As2LuoShuaishuai,XubingSpectroscopicEuZn2As2_11,6EuCuAs_JACS,46EuCuAsnpj,wangLargeTopologicalHall2025,ramMagnetotransportElectronicStructure2024,malickElectronicStructurePhysical2022,jinMultipleMagnetismcontrolledTopological2021,lahaTopologicalHallEffect2021,liManipulationTopologicalPhase2025}. For instance, EuCd$_2$As$_2$ has been proposed as an ideal magnetic Weyl semimetal in its spin-polarized state~\cite{1EuCd2As2_SA,2Soh_IdealWeylsemimetal,32WangLin-Lin_EuCd2As2,3AM_EuCd2As2}, although recent experimental and theoretical studies have challenged this view regarding its nontrivial band topology~\cite{34EuCd2As2Semiconductor,31EuCd2As2ShiYue,38EuCd2As2_SCPMA,33Cuono}. Moreover, multiple members of the Eu$TX$ family, including EuAgAs and EuAuAs, are also known to host topologically nontrivial states~\cite{lahaTopologicalHallEffect2021,jinMultipleMagnetismcontrolledTopological2021,malickElectronicStructurePhysical2022}. These two families share similar crystal symmetries, and certain members exhibit closely matched in-plane lattice constants. This compatibility enables the potential construction of intergrowth structures through alternating stacking of their building units along the crystallographic $c$ axis, sharing common Eu planes~\cite{42Blocklayer}. Using this approach, we successfully designed and synthesized the isostructural intergrowth compound Eu$_2$CuMn$_2$P$_3$ by combining EuCuP and CaAl$_2$Si$_2$-type EuMn$_2$P$_2$~\cite{chen2025}. Furthermore, the synthesis of the phosphide analogue, Eu$_2$CuZn$_2$P$_3$, was reported decades ago and comprehensively characterized recently~\cite{43Eu2123,44Eu2CuZn2P3_PRM}. Thus our design strategy demonstrates high feasibility.

In this work, we hybridize EuCuAs from the Eu$TX$ family and EuZn$_2$As$_2$ from the Eu$M_2X_2$ family to synthesize their intergrowth compound Eu$_2$CuZn$_2$As$_3$. The parent arsenides not only possess closely matched $a$-axis lattice parameters~\cite{tomuschat1981,klufers1980}, but also exhibit nontrivial topological characteristics. EuCuAs undergoes an antiferromagnetic (AFM) transition at 16 K with helical spin ordering~\cite{tong2014,6EuCuAs_JACS,46EuCuAsnpj}. Under applied magnetic fields, a metamagnetic transition induces a noncoplanar spin structure, leading to a pronounced nonlinear anomalous Hall effect (NLAHE) and spin-configuration-dependent topological states~\cite{6EuCuAs_JACS,46EuCuAsnpj}. EuZn$_2$As$_2$ exhibits A-type AFM order below $T_\mathrm{N} = 19.6$ K, where a large NLAHE is also observed, attributed to short-range magnetic correlations~\cite{23EuZn2As2wzc,47EuZn2As2THE}. Pressure-induced topological phase transitions have further been demonstrated in EuZn$_2$As$_2$~\cite{24EuZn2As2LuoShuaishuai}. Their intergrowth compound, Eu$_2$CuZn$_2$As$_3$, was successfully synthesized and characterized, revealing strong coupling between magnetism, charge transport, and band topology. Eu$_2$CuZn$_2$As$_3$ undergoes successive magnetic transitions at $T_\mathrm{M} = 29.3$ K, $T_\mathrm{N} = 19$ K, and $T_\mathrm{SR} = 16.3$ K. In-plane magnetization measurements indicate a field-induced metamagnetic transition. Moreover, we observe strong negative magnetoresistance (nMR) and a pronounced NLAHE. First-principles calculations suggest that the AFM state of Eu$_2$CuZn$_2$As$_3$ potentially features a topologically nontrivial electronic structure. These compelling characteristics demonstrate the efficacy of our exploration strategy for topological materials. Given the high structural compatibility between the Eu$TX$ and Eu$M_{2}X_{2}$ families, the rational application of this design approach promises the discovery of additional magnetic topological materials with intergrowth structures.

\section{METHODS}
\textit{Crystal growth}. Single crystals of Eu$_2$CuZn$_2$As$_3$ were grown using a Sn flux method with high-purity starting materials: Eu ingots (99.9\%), Cu powder (99.9\%), Zn powder (99.99\%), As lumps (99.999\%), and Sn shots (99.99\%). Prior to weighing, the surfaces of the Eu ingots were meticulously cleaned to remove oxide layers and then cut into small pieces. The elements were mixed in a molar ratio of Eu:Cu:Zn:As:Sn = 3:2:2:4:40 under an argon atmosphere in a glove box. The mixture was loaded into an alumina crucible, which was subsequently sealed in an evacuated quartz tube. The growth process involved heating to 600 $^\circ$C over 10 h, maintaining this temperature for 5 h, followed by heating to 1100 $^\circ$C in 10 h with a 36 h dwell time. Finally, the system was cooled to 750 $^\circ$C at a rate of 2 $^\circ$C/h before centrifugation to separate the crystals from the Sn flux.

\textit{Structure determination and composition analysis}. The crystal structure of Eu$_2$CuZn$_2$As$_3$ was determined by single-crystal x-ray diffraction (SCXRD) measurements performed at 150 K using a Bruker D8 Venture diffractometer. The instrument was equipped with an I$\mu$S 3.0 dual-wavelength source (Mo $K\alpha$ radiation, $\lambda$ = 0.71073\,\AA) and an APEX-II CCD detector. Data reduction and correction were performed using the Bruker SAINT software package. Initial structure solution was developed through intrinsic phasing with SHELXT, followed by least-squares refinement using SHELXL2014~\cite{49SHELXL}. Complementary x-ray diffraction measurements were conducted at room temperature using a SmartLab diffractometer (Cu $K\alpha$ radiation) to verify the unit cell parameter $c$. Chemical composition of the single crystals was confirmed by energy-dispersive x-ray spectroscopy (EDS) using an FEI Inspect F50 scanning electron microscope (SEM) equipped with an Ametek EDAX detector.

\textit{Physical property measurements}. Direct-current (dc) magnetization measurements were performed using a Magnetic Property Measurement System (MPMS-3, Quantum Design). Alternating-current (ac) magnetization data were acquired with a Physical Property Measurement System (PPMS DynaCool, Quantum Design) equipped with an ac magnetization measurement system (ACMS-II). Heat capacity was measured by the relaxation-time method using the PPMS. Electrical transport properties were also characterized through standard four-probe measurements on the PPMS platform.

\textit{Theoretical calculations}. First-principles calculations based on density functional theory (DFT) were performed using the Vienna \textit{ab initio} Simulation Package (VASP)~\cite{50abinitio}. The electron-ion interactions were treated with projector augmented wave (PAW) pseudopotentials~\cite{51augmentedwavemethod}, employing a plane-wave cutoff energy of 400\,eV. For the exchange-correlation functional, we adopted the Perdew-Burke-Ernzerhof revised for solids (PBEsol) implementation of the generalized gradient approximation (GGA)~\cite{52J.P.Perdew}. Brillouin zone integration was performed using a $\Gamma$-centered 11$\times$11$\times$2 Monkhorst-Pack $k$-point mesh. To properly account for electron correlation effects, we applied an effective Hubbard $U$ parameter of 5.0\,eV to the Eu 4$f$ orbitals within the Dudarev approach~\cite{53LSDAU}, consistent with previous studies of Eu$^{2+}$ systems~\cite{32WangLin-Lin_EuCd2As2,35EuCd2P2CXY,chen2025}. Structural optimization was carried out by iteratively relaxing both lattice constants and atomic positions until convergence criteria were met: total energy changes below 10$^{-6}$\,eV and Hellmann-Feynman forces smaller than 0.01\,eV/\AA~per atom.
%更新，新的方法

\begin{figure}
	\includegraphics[width=0.45\textwidth]{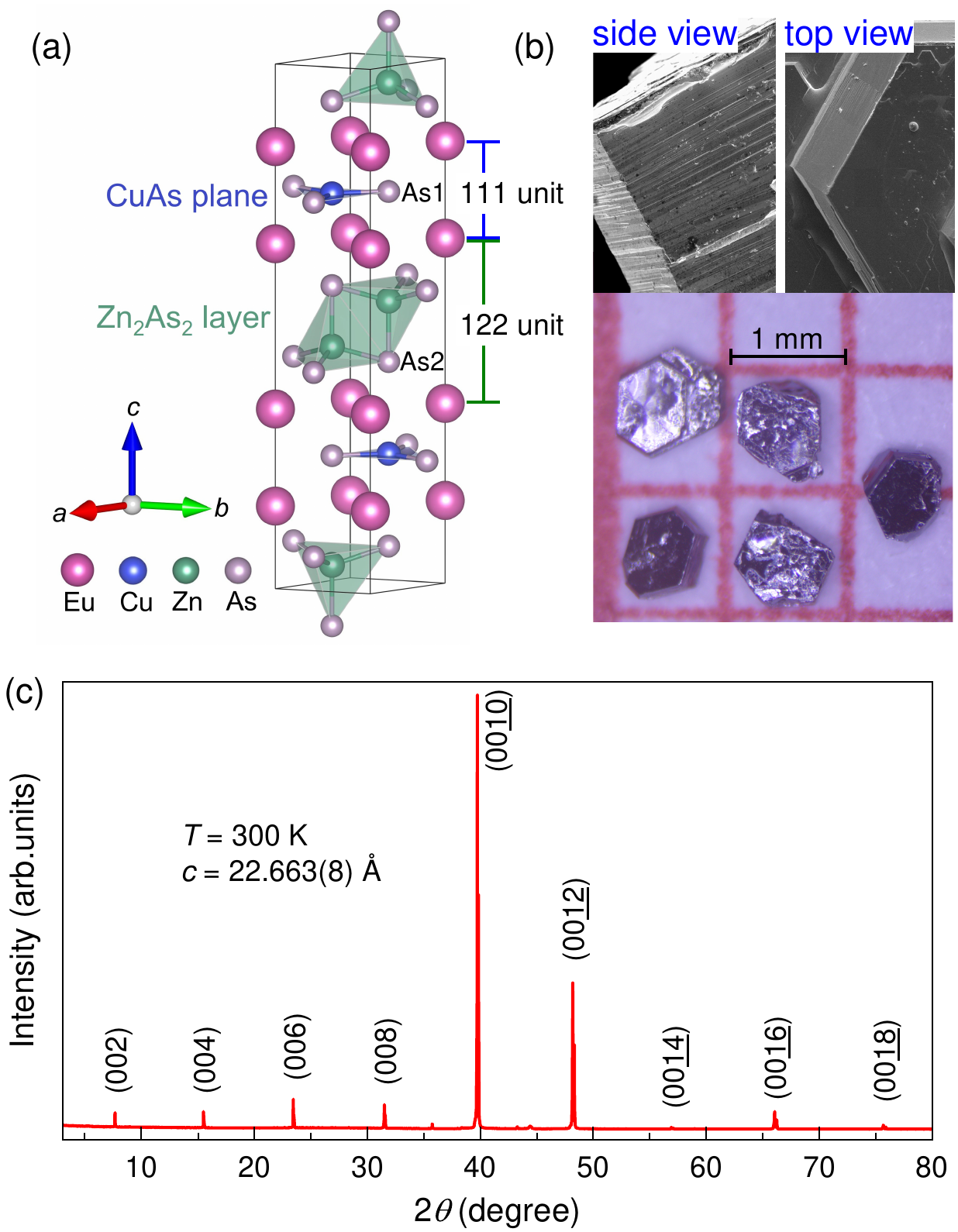}
	\caption{(a) Crystal structure of Eu$_2$CuZn$_2$As$_3$ visualized using VESTA software~\cite{54VESTA}. (b) SEM (top) and optical microscope (bottom) images of Eu$_2$CuZn$_2$As$_3$ single crystals. (c) XRD pattern ($\theta-2\theta$ scan) of the Eu$_2$CuZn$_2$As$_3$ single crystal showing exclusively (00$l$) reflections. The peaks adjacent to the main diffraction peaks are the $K\alpha2$ reflections, while unlabeled weak peaks correspond to unfiltered $K\beta$ reflections.}
	\label{F1}
\end{figure}

\section{RESULTS AND DISSCUSSION}
\subsection{Structrual analysis}

%%%%%%%%%%%%%%%%%%% TABLE1 %%%%%%%%%%%%%%%%%%%%%
\begin{table}
	\caption{Crystallographic data and refinement results for Eu$_2$CuZn$_2$As$_3$ obtained from SCXRD at 150 K~\cite{CCDC}. The occupancies of Eu and As were fixed to 1.0 to avoid unphysical values greater than 1.}
	\begin{ruledtabular}\label{Tab-1}
		\begin{tabular}{ll}	
			Material & Eu$_2$CuZn$_2$As$_3$ \\
			\hline
			Crystal system  & Hexagonal \\	
			Space group & $P6_3/mmc$ (No. 194)  \\   
			$a$ (\AA) & 4.2272(2) \\                               
			$c$ (\AA) & 22.4932(14) \\
			%			$\alpha$ = $\beta$  & 90$\rm^o$ \\
			%			$\gamma$ & 120$\rm^o$   \\    
			$V$ (\AA$^3$) & 348.09(4) \\			                 
			$Z$ & 2 \\	
			$\rho_\mathrm{calc}$ (g/cm$^3$)  & 6.864      \\ 		
			Temperature (K) & 150 \\
			Radiation & Mo $K\alpha$ \\
			Reflections collected & 3559 \\
			Independent reflections & 176  \\
			$R\rm_{int}$ & 0.0969  \\
			Goodness-of-fit & 1.361 \\
			$R_1 [F^2>2\sigma(F^2)]$\footnote{$R_1=\Sigma||F_o|-|F_c||/\Sigma|F_o|$.} & 0.0342 \\
			$wR_2 [F^2]$\footnote{$wR_2=[\Sigma w(F_o^2-F_c^2)^2/\Sigma w(F_o^2)^2]^{1/2}$.}  & 0.0875 \\	
			Fractional coordinates & \\
			Eu & (0,0,0.15754(5))   \\                            
			Cu & (1/3,2/3,1/4) \\
			Zn & (1/3,2/3,0.54157(11))  \\
			As1 (CuAs plane) & (1/3,2/3,3/4)  \\			
			As2 (Zn$_2$As$_2$ layer) & (1/3,2/3,0.07412(10)) \\
			Site occupancy & \\
			Eu & 1.0   \\                            
			Cu & 0.975(13) \\
			Zn & 0.984(16)  \\
			As & 1.0  \\
		\end{tabular}
	\end{ruledtabular}
\end{table}
%Figure \ref{F1}(a) depicts the crystal structure of Eu$_2$CuZn$_2$As$_3$ in a $P6_3/mmc$ space group, characterized by a periodic stacking of triangular Eu$^{2+}$ layer, honeycomb [CuAs]$^{2-}$ plane, Eu$^{2+}$ layer, and [Zn$_2$As$_2$]$^{2-}$ layer comprising of edge-sharing ZnAs$_4$ tetrahedra along the crystallographic $c$ axis. The structure of Eu$_2$CuZn$_2$As$_3$ can be regarded as a hybrid of EuCuAs and EuZn$_2$As$_2$. In such heterogeneous structure, competing interactions are commonly observed. For instance, the magnetic coupling between the neighbor Eu layer in Eu$_2$CuZn$_2$As$_3$ could be mediated by either the CuAs planes or the Zn$_2$As$_2$ layers. These complex magnetic interactions give rise to the unique magnetism observed in Eu$_2$CuZn$_2$As$_3$, which will be elaborated upon subsequently. 

Figure~\ref{F1}(a) shows the crystal structure of Eu$_2$CuZn$_2$As$_3$ in the $P6_3/mmc$ space group. The structure consists of periodic stacking layers along the crystallographic $c$-axis: triangular Eu$^{2+}$ planes, honeycomb [CuAs]$^{2-}$ planes, and [Zn$_2$As$_2$]$^{2-}$ layers composed of edge-sharing ZnAs$_4$ tetrahedra. Eu$_2$CuZn$_2$As$_3$ can be viewed as a structural hybrid of EuCuAs and EuZn$_2$As$_2$, where complex interactions typically emerge. In particular, the magnetic coupling between adjacent Eu$^{2+}$ planes may be mediated either through the [CuAs]$^{2-}$ planes or the [Zn$_2$As$_2$]$^{2-}$ layers. These complex magnetic interactions are responsible for the unique magnetic properties observed in Eu$_2$CuZn$_2$As$_3$, as will be discussed in detail later.

% any given Eu atom is influenced by two distinct interlayer magnetic interactions: one mediated by CuAs planes and the other one by Zn$_2$As$_2$ layers.

%The CuAs planes and Zn$_2$As$_2$ layers are derived from EuCuAs and EuZn$_2$As$_2$, respectively, thus Eu$_2$CuZn$_2$As$_3$ can be regarded as a natural heterojunction. 

The synthesis of Eu$_2$CuZn$_2$As$_3$ was achieved through intentional design rather than serendipitous discovery. The formation of this phase was enabled by the closely matched lattice parameter $a$ and similar crystal symmetry between EuCuAs and EuZn$_2$As$_2$~\cite{tomuschat1981,klufers1980}. This rationale similarly explains the synthesis of previously reported isostructural compounds Ca$_2$CuZn$_2$P$_3$ and Eu$_2$CuZn$_2$P$_3$~\cite{43Eu2123}, as well as the recently synthesized Eu$_2$CuMn$_2$P$_3$~\cite{chen2025}. EuCuAs ($Z$ = 2) crystallizes in the hexagonal SrPtSb-type structure (space group $P6_3/mmc$, No. 194), with lattice parameters $a_{\mathrm{111}}$ = 4.2598(3)\,\AA\ and $c_{\mathrm{111}}$ = 8.2857(9)\,\AA~\cite{tong2014}. EuZn$_2$As$_2$ ($Z$ = 1) adopts the trigonal CaAl$_2$Si$_2$-type structure (space group $P\bar{3}m1$, No. 164), characterized by lattice parameters $a_{\mathrm{122}}$ = 4.21118(3)\,\AA\ and $c_{\mathrm{122}}$ = 7.18114(6)\,\AA~\cite{23EuZn2As2wzc}. The difference in lattice parameter $a$ between these compounds is marginal, being only approximately 1\%. Since Eu$_2$CuZn$_2$As$_3$ forms through sharing a Eu atomic layer between half a unit cell of EuCuAs ($Z$ = 2) and one unit cell of EuZn$_2$As$_2$, we anticipate that its lattice parameter $a_{\mathrm{2123}}$ should lie between $a_{\mathrm{111}}$ and $a_{\mathrm{122}}$, while $c_{\mathrm{2123}}$ should approximately equal $2\times(\frac{1}{2}c_{\mathrm{111}} + c_{\mathrm{122}})$. Our predictions for the lattice constants of Eu$_2$CuZn$_2$As$_3$ were confirmed by SCXRD measurements. Table~\ref{Tab-1} presents the crystallographic parameters and refinement results obtained from SCXRD data for Eu$_2$CuZn$_2$As$_3$ at 150\,K.

%The optical picture and the SEM image of Eu$_2$CuZn$_2$As$_3$ single crystals are shown in Fig. \ref{F1}(b). The typical dimension of the crystal is $\sim$0.5$\times$0.5$\times$0.4 mm. Furthermore, the SEM image clear shows unsmooth side facet of Eu$_2$CuZn$_2$As$_3$ single crystal, revealing its layer-by-layer crystal growth process, consistent with the complex intergrowth structure of Eu$_2$CuZn$_2$As$_3$. To evaluate the crystal quality and determine the $c_{\mathrm{2123}}$ of Eu$_2$CuZn$_2$As$_3$ at room temperature, we performed $\theta-2\theta$ XRD measurements on single crystals, as shown in Fig.~\ref{F1}(c). A series of sharp (00$l$) diffraction peaks were observed, implying the high quality of Eu$_2$CuZn$_2$As$_3$ single crystals. The derived lattice parameter $c_{\mathrm{2123}}$ = 22.663(8)\,\AA, calculated from the (00$l$) reflections, agrees well with the predicted value of $c_{\mathrm{111}} + 2\times c_{\mathrm{122}} = 22.648$\,\AA. We note that the room-temperature $c_{\mathrm{2123}}$ value shows a slight deviation from that obtained by SCXRD at 150\,K, which can be attributed to thermal contraction effects on the lattice parameters. These findings collectively suggest that the high quality crystals of intergowth Eu$_2$CuZn$_2$As$_3$ were successfully synthesized.

Figure~\ref{F1}(b) displays optical and SEM images of Eu$_2$CuZn$_2$As$_3$ single crystals, with typical dimensions of approximately 0.7$\times$0.7$\times$0.4\,mm$^3$. The SEM image clearly reveals uneven side facets, indicating a layer-by-layer crystal growth process, which is consistent with the characteristic stacking of block layers along the $c$ axis in Eu$_2$CuZn$_2$As$_3$. To assess crystal quality and determine the room-temperature lattice parameter $c_{\mathrm{2123}}$, we conducted $\theta-2\theta$ XRD measurements, as shown in Fig.~\ref{F1}(c). The observation of sharp (00$l$) diffraction peaks confirms the high crystalline quality of the Eu$_2$CuZn$_2$As$_3$ single crystals. The derived lattice parameter $c_{\mathrm{2123}} = 22.663(8)$\,\AA, calculated from the (00$l$) reflections, shows excellent agreement with the predicted value of $c_{\mathrm{111}} + 2\times  c_{\mathrm{122}} = 22.648$\,\AA. The slight discrepancy between this room-temperature value and that obtained from SCXRD at 150\,K can be attributed to thermal contraction effects. These results collectively demonstrate the successful synthesis of high-quality intergrowth Eu$_2$CuZn$_2$As$_3$ single crystals.

%The synthesis of Eu$_2$CuZn$_2$As$_3$ is not discovered by accident, but by intentionally design. The formation of the phase Eu$_2$CuZn$_2$As$_3$ was facilitated by the closely matched lattice parameter $a$ and similar crystal symmetry between EuCuAs and EuZn$_2$As$_2$. Similarly, the decades ago reported isostructural Ca$_2$CuZn$_2$P$_3$ and Eu$_2$CuZn$_2$P$_3$, as well as the recently synthesized Eu$_2$CuMn$_2$P$_3$, were synthesized for the same reason. EuCuAs ($Z$ = 2) crystallizes in the hexagonal SrPtSb-type structure (space group $P6_3/mmc$, No. 194), whose cell parameters $a_{111}$ = 4.2598(3) \AA\ and $c_{111}$ = 8.2857(9) \AA. EuZn$_2$As$_2$ ($Z$ = 1) crystallizes in the trigonal CaAl$_2$Si$_2$-type structure with a $P\bar3m1$ space group (No. 164), whose lattice parameter $a_{122}$ = 4.21118(3) \AA\ and $c_{122}$ = 7.18114(6) \AA. The different in the lattice parameter $a$ between the two compounds is marginal, only about 1\%.  Since Eu$_2$CuZn$_2$As$_3$ is formed by sharing an Eu atom layer between half unit cell of EuCuAs ($Z$ = 2) and one unit cell of EuZn$_2$As$_2$, it is anticipated that the lattice parameter $a_{2123}$ of Eu$_2$CuZn$_2$As$_3$ will fall between those of $a_{111}$ \AA\ and $a_{122}$ \AA, while $c_{2123}$ should be approximately equal to the sum of $\frac{1}{2}c_{111}$ and $c_{122}$. Our assumption of the lattice constants for Eu$_2$CuZn$_2$As$_3$ is confirmed by the SCXRD. Table \ref{Tab-1} lists the crystallographic parameters and the refinement results of SCXRD data for Eu$_2$CuZn$_2$As$_3$ at 150 K.

\subsection{Magnetism}

\begin{figure*}
	\includegraphics[width=\textwidth]{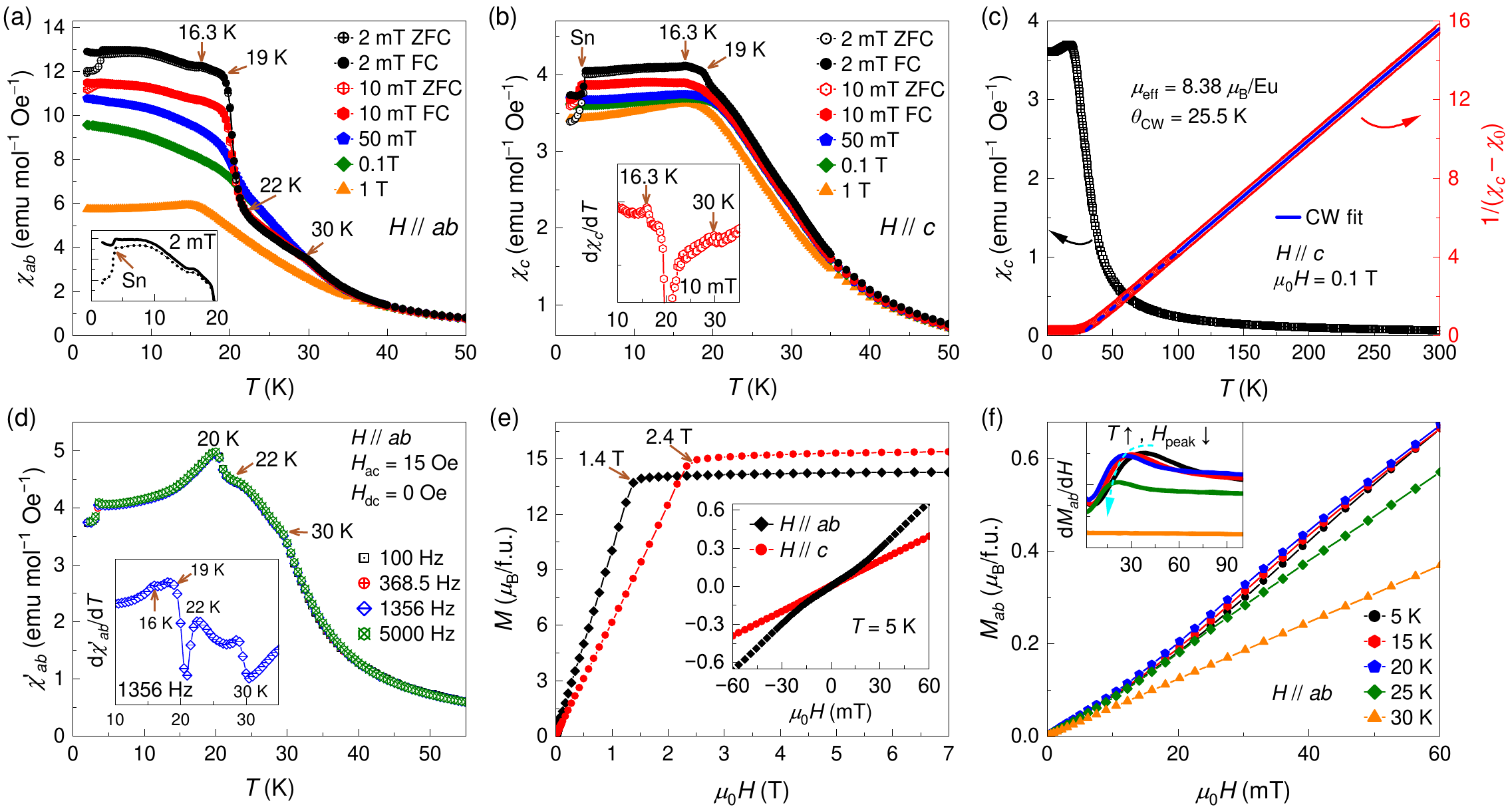}
	\caption{(a) Temperature-dependent in-plane magnetic susceptibility $\chi_{ab}(T)$ of Eu$_2$CuZn$_2$As$_3$ under different magnetic fields ($H\parallel ab$). The inset shows the splitting between ZFC (dash line) and FC (solid line) curves below 19\,K under a 2\,mT field. (b) Out-of-plane magnetic susceptibility $\chi_c(T)$ under various applied fields ($H\parallel c$). The inset shows the temperature derivative $d\chi_c/dT$, highlighting the magnetic transitions in Eu$_2$CuZn$_2$As$_3$. (c) $\chi_c(T)$ measured at 0.1 T (black, left axis) with corresponding CW analysis for the 100--300 K temperature range (right axis). (d) In-phase component ($\chi^{\prime}_{ab}$) of the ac magnetic susceptibility ($H\parallel ab$) at various frequencies. The inset presents $d\chi^{\prime}_{ab}/dT$ at 1356 Hz. (e) Field-dependent magnetization $M(H)$ at 5 K for both in-plane (black diamonds) and out-of-plane (red dots) field orientations. The inset displays the magnetization $M(H)$ in the low-field range ($\mu_0H \leq 60$~mT). (f) In-plane magnetization curves $M_{ab}(H)$ measured under small applied fields at various temperatures. The inset shows the field derivative $dM_{ab}/dH$.}
	\label{F2}
\end{figure*}
%(a)加的inset

%Moreover, both compounds features a low carrier concentration on the order of $10^{20}$, and could exhibit a strong negative MR effect upon applying the field.

%As previously mentioned, the structural units of Eu$_2$CuZn$_2$As$_3$ are derived from EuCuAs and EuZn$_2$As$_2$, thus featuring the magnetic characteristics similar to those of EuCuAs and EuZn$_2$As$_2$. EuCuAs was reported to be an antiferromagnetic (AFM) topological insulator with a helical spin order below $T_\mathrm{N}\sim$ 16 K, while EuZn$_2$As$_2$ was an A-type antiferromagnet with a transition temperature $T_\mathrm{N}$ = 19.6 K, and a topological phase transition is believed to be induced by pressure. Consequently, Eu$_2$CuZn$_2$As$_3$ is found to exhibit properties similar to those of EuCuAs and EuZn$_2$As$_2$. Eu$_2$CuZn$_2$As$_3$ is also an antiferromagnet, exhibiting transitions around 16 K and 19 K, originating from the magnetic couplings within parent units. However, the magnetic structure and in-plane magnetic behavior of Eu$_2$CuZn$_2$As$_3$  is better complicated than EuCuAs and EuZn$_2$As$_2$ due to the emergent magnetic interactions.

As previously mentioned, the structural blocks of Eu$_2$CuZn$_2$As$_3$ derive from the parent compounds EuCuAs and EuZn$_2$As$_2$. This structural inheritance therefore results in magnetic characteristics for Eu$_2$CuZn$_2$As$_3$ similar to those observed in its parent systems. EuCuAs has been identified as an AFM topological insulator or Weyl semimetal that develops helical spin ordering below its N\'{e}el temperature ($T_\mathrm{N} \sim 16$\,K)~\cite{tong2014,45EuCuAsAndrew,46EuCuAsnpj,6EuCuAs_JACS}, while EuZn$_2$As$_2$ exhibits A-type antiferromagnetism with $T_\mathrm{N} = 19.6$\,K~\cite{23EuZn2As2wzc,blawat2022,47EuZn2As2THE}, where pressure-induced topological phase transitions have been demonstrated~\cite{24EuZn2As2LuoShuaishuai}. Our investigations confirm that Eu$_2$CuZn$_2$As$_3$ indeed manifests magnetic properties analogous to its parent compounds, showing clear transitions near 16\,K and 19\,K that correspond to the characteristic temperatures of its constituent structural units. Notably, Eu$_2$CuZn$_2$As$_3$ displays more complex magnetic ground states and in-plane magnetic behaviors compared to its parent compounds, arising from emergent magnetic interactions between its hybridized structural units.

The magnetic properties of Eu$_2$CuZn$_2$As$_3$ are summarized in Fig.~\ref{F2}. Figure~\ref{F2}(a) displays the in-plane magnetic susceptibility $\chi_{ab}(T)$, which exhibits multiple features associated with magnetic phase transitions. These include: (i) a weak hump near 30\,K in low-field $\chi_{ab}(T)$ curves, (ii) a pronounced increase below 22\,K, (iii) a distinct inflection point at approximately 19\,K, and (iv) a subtle kink around 16.3\,K. Similar features appear in both the out-of-plane susceptibility $\chi_c(T)$ and the in-phase component $\chi^{\prime}_{ab}(T)$ of the ac magnetic susceptibility with $H\parallel ab$, although the 30\,K hump and the susceptibility enhancement below 22\,K are less pronounced, as shown in Figs.~\ref{F2}(b) and~\ref{F2}(d). Notably, their temperature derivatives $d\chi_c/dT$ and $d\chi^{\prime}_{ab}/dT$ all exhibit clear peaks at these characteristic temperatures (see insets). These magnetic signatures are systematically suppressed with increasing magnetic field. Furthermore, the ZFC and FC measurements for both $\chi_{ab}(T)$ and $\chi_c(T)$ reveal a small bifurcation below 19\,K under weak fields (2\,mT) [Figs.~\ref{F2}(a) and~\ref{F2}(b)], suggesting a minor ferromagnetic (FM) component that may originate from canted or uncompensated spins in the Eu$^{2+}$ sublattice. This weak ferromagnetism could be particularly significant, as it suggests that exotic phenomena might be achievable through the manipulation of magnetic domains with small external fields, resembling the situation observed in the AFM topological material Mn$_3$Sn~\cite{kimata2019,kuroda2017}. In addition, the slight drop below 5\,K observed in both the low-field $\chi_{ab}(T)$ and $\chi_c(T)$ curves is caused by a superconductivity signal originating from a tiny amount of residual Sn flux on the sample surface.

The first magnetic transition observed near 30\,K, designated as $T_\mathrm{M}$, is corroborated by a weak peak in the specific heat data at 29.3\,K, as shown in Fig.~\ref{F3}(a). The marginal signature of this transition in the $\chi_c(T)$ [inset of Fig.~\ref{F2}(b)], combined with the limited magnitude of the $C_\mathrm{p}(T)$ peak at 29.3\,K, suggests that the transition at $T_\mathrm{M}$ corresponds to magnetic ordering confined within the $ab$-plane, indicative of quasi-two-dimensional (quasi-2D) behavior. Below 22\,K, the gradual increase in susceptibility followed by an inflection point near 19\,K signals the emergence of long-range magnetic ordering in the Eu$^{2+}$ sublattice. This interpretation is supported by a corresponding peak in $C_\mathrm{p}(T)$ at 18\,K. We therefore identify 22\,K as the onset temperature and 19\,K as the N\'{e}el temperature ($T_\mathrm{N}$) for this transition, with further discussion provided in the heat capacity section. The feature observed at 16.3\,K, labeled as $T_\mathrm{SR}$, is tentatively attributed to a spin reorientation transition within the Eu$^{2+}$ sublattice, while no corresponding anomaly is observed in the specific heat data at this temperature.

The observation of multiple magnetic transitions in Eu$_2$CuZn$_2$As$_3$ is consistent with its intergrowth crystal structure. The differing thicknesses of the constituent EuCuAs and EuZn$_2$As$_2$ blocks, along with distinct superexchange pathways within these structural units, naturally lead to varying magnetic coupling strengths between Eu planes, thereby inducing successive magnetic ordering transitions. The transition temperatures $T_\mathrm{N} = 19$\,K and $T_\mathrm{SR} = 16.3$\,K closely match the N\'{e}el temperatures of EuZn$_2$As$_2$ and EuCuAs, respectively~\cite{23EuZn2As2wzc,6EuCuAs_JACS}. This correspondence suggests that these transitions are primarily governed by exchange correlations mediated through the Zn$_2$As$_2$ layers and CuAs planes. The spin reorientation at $T_\mathrm{SR}$ likely stems from the competition between exchange interactions and magnetic-dipole interactions at low temperatures. The latter may play a significant role in shaping the spin structure of Eu-based magnetic systems due to the zero orbital angular momentum and large ordered moment of the Eu$^{2+}$ ions, as previously discussed in studies on EuZn$_2$P$_2$ and other systems~\cite{19Berry_EuZn2P2,johnston2016}. Similar spin reorientation transitions have also been observed in other Eu-based materials, such as EuMn$_2$As$_2$ and Eu$_2$CuMn$_2$P$_3$~\cite{anand2016,chen2025}. Recent studies of isostructural phosphide analogue Eu$_2$CuZn$_2$P$_3$ revealed a weak susceptibility anomaly near 26\,K (approaching the N\'{e}el temperature of EuZn$_2$P$_2$) under low field conditions, analogous to the spin reorientation characteristics observed in our Eu$_2$CuZn$_2$As$_3$ sample~\cite{44Eu2CuZn2P3_PRM,19Berry_EuZn2P2}. Notably, the quasi-2D transition at $T_\mathrm{M} \sim 30$\,K represents an emergent phenomenon absent in both parent compounds. In fact, Eu$_2$CuZn$_2$P$_3$ also exhibits enhanced AFM ordering at 40\,K, significantly exceeding the ordering temperatures of its parent compounds EuZn$_2$P$_2$ (23.5\,K) and EuCuP (30\,K)~\cite{44Eu2CuZn2P3_PRM,19Berry_EuZn2P2,45EuCuAsAndrew}. The enhancement of magnetic interactions in these intergrowth systems may originate from the interfacial coupling between stacked block layers, which warrants deeper investigation.

Figure~\ref{F2}(e) presents the $M_{ab}(H)$ (in-plane) and $M_c(H)$ (out-of-plane) magnetization curves measured at 5\,K. The in-plane saturation field of 1.4\,T is notably lower than the out-of-plane value of 2.4\,T. Crucially, the $M_{ab}(H)$ curve exhibits a distinct positive curvature in the low-field regime ($<30$\,mT), a feature absent in the $M_c(H)$ response. The substantially smaller in-plane saturation field, together with the metamagnetic transition observed under $H\parallel ab$, indicates an easy-plane AFM configuration in Eu$_2$CuZn$_2$As$_3$. 
Detailed analysis of the low-field $M_{ab}(H)$ curves [Fig.~\ref{F2}(f)] demonstrates that this positive curvature emerges below the transition at $T_\mathrm{M} = 30$\,K. The field derivative $dM_{ab}/dH$ plotted in the inset reveal broad peaks corresponding to the maximum slopes of the $M_{ab}(H)$ curves, with the peak field $H_\mathrm{peak}$ systematically shifting to lower values with increasing temperature. This characteristic nonlinearity in the $M_{ab}(H)$ data is commonly observed in Eu- and Gd-based layered compounds with A-type AFM structures, including trigonal EuMg$_2$Bi$_2$ and EuAl$_2$Ge$_2$, as well as tetragonal EuGa$_4$ and GdRh$_2$Si$_2$~\cite{pakhira2023b,pakhira2023,slade2025}. Moreover, similar nonlinear feature is observed in the isostructural Eu$_2$CuZn$_2$P$_3$~\cite{44Eu2CuZn2P3_PRM}. Previous studies have shown that this behavior can arise from magnetic-field-induced moment reorientation within AFM domains~\cite{pakhira2023b,pakhira2023,slade2025}.

%In the case of Eu$_2$CuZn$_2$As$_3$, this metamagnetic magnetic transition under low fields should be linked to the EuCuAs block. A helical spin order was well established for the parent compound EuCuAs, while a Eu moments reorientation within the $ab$ plane from the helical to a canted double-period AFM structure can be induced by a critical field of 0.3 T. In contrast, the ground state of EuZn$_2$As$_2$ is the simple A-type AFM, and such metamagnetic transition is absent in EuZn$_2$As$_2$. Notably, the critical field for the metamagnetic transition in Eu$_2$CuZn$_2$As$_3$, i.e. the value of $H_{peak}$, is only $\sim$38 mT at 5 K, which is significantly lower than that of EuCuAs. This substantial disparity  implies the much weaker inter-block magnetic coupling between the EuCuAs blocks, due to the intercalation of the EuZn$_2$As$_2$ blocks, which leads to the considerably more unstable spin state under magnetic fields.

The low-field metamagnetic transition in Eu$_2$CuZn$_2$As$_3$ originates primarily from the EuCuAs structural blocks. This assignment is supported by comparative analysis with the parent compounds: EuCuAs exhibits a well-documented helical spin order that undergoes field-induced reorientation to a canted double-period AFM structure at a critical field of 0.3\,T~\cite{46EuCuAsnpj}, whereas EuZn$_2$As$_2$ adopts a simple A-type AFM ground state without the metamagnetic transition~\cite{23EuZn2As2wzc}. Remarkably, the critical field ($H_\mathrm{peak} \sim 38$\,mT at 5\,K) for Eu$_2$CuZn$_2$As$_3$ is nearly an order of magnitude lower than that of EuCuAs. This dramatic reduction indicates significantly weakened inter-block magnetic coupling between EuCuAs units, a direct consequence of the intervening EuZn$_2$As$_2$ blocks that enhance spin-state instability under applied magnetic fields.

%Figure \ref{F2}(c) presents the $\chi_c$ under a measured field of 0.1 T and the corresponding Curie-Weiss analysis. The Curie-Weiss law $\chi$ = $\chi_0$ + $C/(T-\theta_\mathrm{CW})$ is used to fit the $\chi_c$ data from 100 to 300 K. The fitting yields Weiss temperature $\theta_\mathrm{CW}$ = 25.5 K, the effective moment $\mu_\mathrm{eff}$ = 8.37 $\mu_\mathrm{B}$/Eu (which is derived from a Curie constant of $C$ = 17.5 emu K mol$^{-1}$ Oe$^{-1}$ f.u.$^{-1}$), and a negligible temperature-independent $\chi_0$ = 1.55 $\times$ 10$^{-4}$ emu mol$^{-1}$ Oe$^{-1}$. The $\mu_\mathrm{eff}$ of Eu$_2$CuZn$_2$As$_3$ is close to the theoretical magnetic moment of 7.94 $\mu_\mathrm{B}$ for Eu$^{2+}$. The positive $\theta_\mathrm{CW}$ was observed in many Eu-based layered materials, such as EuZn$_2$As$_2$ and EuCuAs, resulting from the dominant FM interactions within Eu planes. The Eu$^{2+}$ oxidation state is also corroborated by the saturated moments in Fig.  \ref{F2}(e), which are close to theorectical value of 14$\mu_\mathrm{B}$.

Figure~\ref{F2}(c) displays the out-of-plane magnetic susceptibility ($\chi_c$) measured at 0.1\,T, along with the corresponding Curie-Weiss (CW) analysis. The CW law, expressed as	$\chi = \chi_0 + {C}/{(T - \theta_\mathrm{CW})}$, was applied to fit the $\chi_c$ data between 100 and 300\,K. The fit yields a Weiss temperature $\theta_\mathrm{CW} = 25.5$\,K, an effective moment $\mu_\mathrm{eff} = 8.37\,\mu_\mathrm{B}$/Eu (calculated from the Curie constant $C = 17.5$\,emu\,K\,mol$^{-1}$\,Oe$^{-1}$\,f.u.$^{-1}$), and a negligible temperature-independent term $\chi_0 = 1.55 \times 10^{-4}$\,emu\,mol$^{-1}$\,Oe$^{-1}$. The obtained $\mu_\mathrm{eff}$ agrees well with the theoretical value of $7.94\,\mu_\mathrm{B}$ for free Eu$^{2+}$ ions. The positive $\theta_\mathrm{CW}$, commonly observed in Eu-based layered materials such as EuZn$_2$As$_2$ and EuCuAs, results from dominant FM interactions within the Eu planes~\cite{23EuZn2As2wzc,6EuCuAs_JACS}. Furthermore, the Eu$^{2+}$ oxidation state is corroborated by the saturation magnetization shown in Fig.~\ref{F2}(e), where the measured moment approaches the theoretical $14\,\mu_\mathrm{B}$ value for two Eu$^{2+}$.

\subsection{Heat capacity}

\begin{figure*}
	\includegraphics[width=0.75\textwidth]{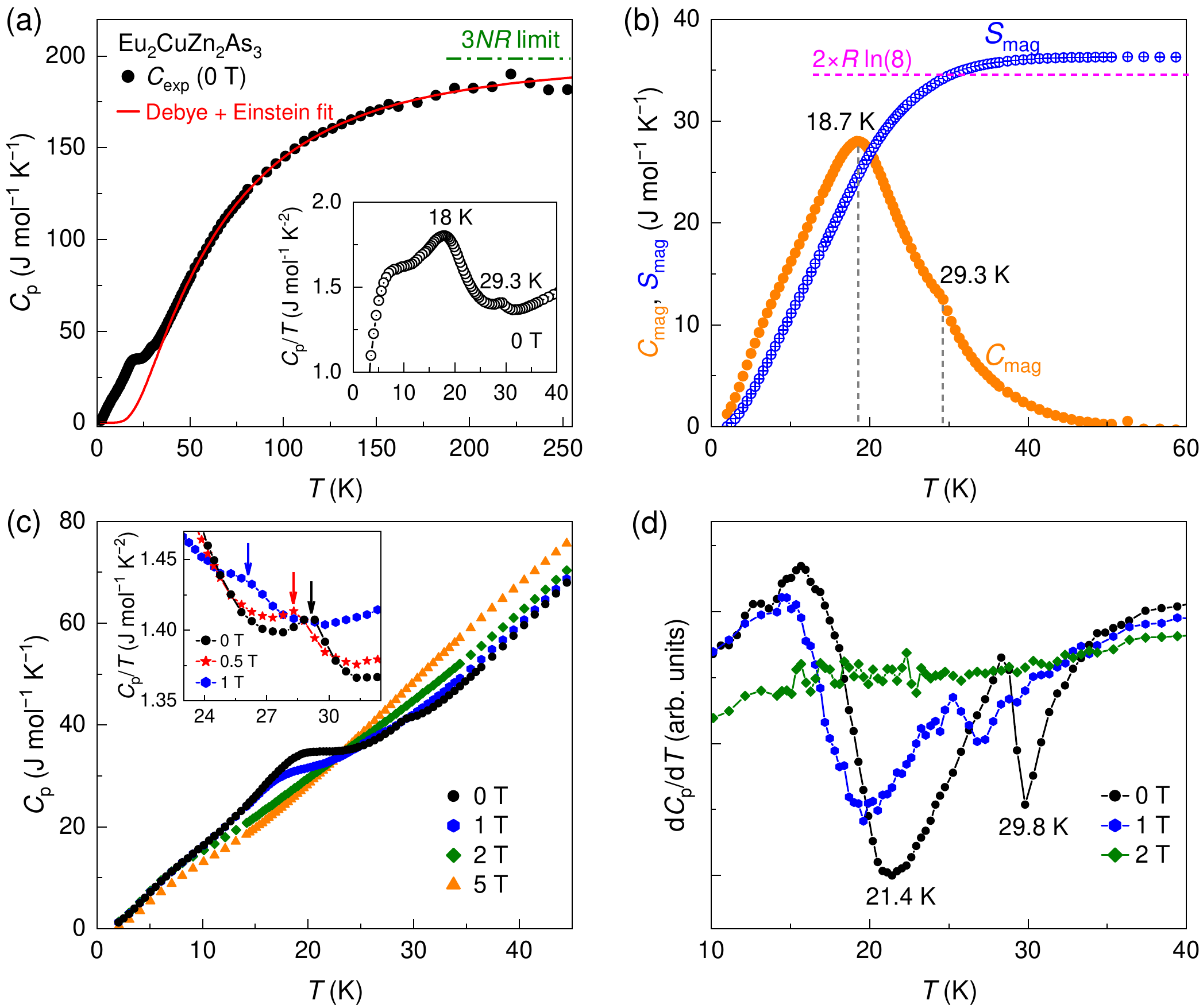}
	\caption{(a) Temperature dependence of the zero-field heat capacity $C_\mathrm{p}(T)$ of Eu$_2$CuZn$_2$As$_3$. The red curve represents a fit combining the Debye and Einstein models in the temperature range 50--190 K. The inset plots $C_\mathrm{p}/T$ versus $T$ below 40 K. (b) Temperature dependence of the magnetic specific heat $C_\mathrm{mag}(T)$ (orange) and the derived magnetic entropy $S_\mathrm{mag}(T)$ (blue). (c) $C_\mathrm{p}(T)$ at various applied magnetic fields. The inset plots $C_\mathrm{p}/T$ versus $T$ at 0 T (black), 0.5 T (red), and 1 T (blue). (d) Temperature derivative of the heat capacity, $dC_\mathrm{p}/dT$, in the range 10--40 K.}
	\label{F3}
\end{figure*}

To gain deeper insight into the phase transitions of Eu$_2$CuZn$_2$As$_3$, we measured the specific heat ($C_\mathrm{p}$) of single-crystal samples under various fields, as shown in Fig.~\ref{F3}. The zero-field $C_\mathrm{p}(T)$ data in Fig.~\ref{F3}(a) reveal two anomalies: a broad peak at 18\,K and a small kink at 29.3\,K. These anomalies are highlighted in the $C_\mathrm{p}/T$ vs. $T$ plot shown in the inset of Fig.~\ref{F3}(a), corresponding to the transition temperatures $T_\mathrm{N} = 19$\,K and $T_\mathrm{M} = 30$\,K observed in magnetic susceptibility measurements. The broad shoulder near 8\,K, commonly seen in $S = 7/2$ magnets, can be explained within the framework of molecular field theory~\cite{johnston2015}.

To isolate the magnetic contribution to the heat capacity, we fitted the zero-field $C_\mathrm{p}(T)$ between 50 and 190\,K using the linear combination of Debye and Einstein models. The lattice specific heat is expressed as:
\begin{equation}
	\begin{split}
		C_\mathrm{lattice}(T) = 9\alpha NR(\frac{T}{\Theta_\mathrm{D}})^3 \int_{0}^{\Theta_\mathrm{D}/T}\frac{x^4 e^x}{e^x - 1}dx \\
		+ 3(1 - \alpha)NR(\frac{\Theta_\mathrm{E}}{T})^2  \frac{e^{\Theta_\mathrm{E}}/T}{(e^{\Theta_\mathrm{E}} - 1)^2},
	\end{split}	
\end{equation}
where the first and second terms represent the Debye and Einstein contributions, respectively. Here $\Theta_\mathrm{D}$, $\Theta_\mathrm{E}$, and $\alpha$ denote the Debye temperature, Einstein temperature, and fractional weight of the Debye component, respectively. The optimized parameters are $\Theta_\mathrm{D} = 377.4$\,K, $\Theta_\mathrm{E} = 133.0$\,K, and $\alpha = 0.42$, with the fitted curve shown as the red solid line in Fig.~\ref{F3}(a).
	
The magnetic specific heat $C_\mathrm{mag}(T)$, obtained by subtracting the lattice contribution, is presented in Fig.~\ref{F3}(b). $C_\mathrm{mag}(T)$ exhibits two distinct magnetic transitions at 18.7\,K and 29.3\,K, but shows no resolvable anomaly near $T_\mathrm{SR} = 16.3$\,K. This absence of a specific heat feature at $T_\mathrm{SR}$ supports our interpretation that the susceptibility kinks at 16.3\,K correspond to a spin reorientation transition, which contributes negligibly to the heat capacity. The magnetic entropy $S_\mathrm{mag}$ in Fig.~\ref{F3}(b) was calculated via:
\begin{equation}
	S_\mathrm{mag}(T) = \int^{T}_{0} \frac{C_\mathrm{mag}(T')}{T'} dT'.
\end{equation}
The saturation value $S_\mathrm{mag} \approx 36.8$\,J\,mol$^{-1}$\,K$^{-1}$ slightly exceeds the theoretical $S_\mathrm{mag} = 2R\ln(2S + 1) = 2R\ln(8)$ for Eu$^{2+}$ ($S=7/2$), with this minor discrepancy attributed to fitting uncertainties.

%两磁相变
%40 K以下显著磁贡献

Figure~\ref{F3}(c) presents $C_\mathrm{p}(T)$ for Eu$_2$CuZn$_2$As$_3$ under several fields. The broad transition peak corresponding to $T_\mathrm{N}$ systematically shifts to lower temperatures with increasing field. Furthermore, the anomaly at $T_\mathrm{M}$ is progressively suppressed under magnetic fields, as clearly shown in the zoomed-in $C_\mathrm{p}/T$ versus $T$ plot in the inset. The field-induced suppression of both transition temperatures is further evidenced by the derivative $dC_\mathrm{p}/dT$ data presented in Fig.~\ref{F3}(d). Notably, at 2\,T, both transition features become completely indistinguishable in the $dC_\mathrm{p}/dT$ curves. This systematic reduction of transition temperatures under applied magnetic fields unambiguously confirms the AFM nature of the transitions at both $T_\mathrm{N}$ and $T_\mathrm{M}$.

Notably, neither transition exhibits a characteristic $\lambda$-type peak typically associated with AFM ordering. The transition at $T_{\mathrm{M}} = 29.3$\,K manifests as only a subtle feature in the $C_{\mathrm{mag}}(T)$ curve [Fig.~\ref{F3}(b)]. However, the significant rise in $C_\mathrm{mag}$ below 40\,K indicates substantial contributions from short-range magnetic correlations. These observations support the quasi-2D nature of the magnetic behavior near $T_{\mathrm{M}}$, reflecting spin ordering confined to the $ab$-plane. In contrast, the three-dimensional ordering transition at $T_{\mathrm{N}}$ displays an unconventional plateau-like feature below 22\,K in $C_{\mathrm{p}}(T)$, distinct from the sharp peaks observed in both parent compounds EuCuAs and EuZn$_2$As$_2$, other Eu-based 111/122 materials, and the isostructural phosphide analogue Eu$_2$CuZn$_2$P$_3$~\cite{6EuCuAs_JACS,23EuZn2As2wzc,19Berry_EuZn2P2,28EuCd2As2npj,5EuCd2P2_AM,pakhira2023b,44Eu2CuZn2P3_PRM}. This anomalous behavior originates from the intergrowth structure of Eu$_2$CuZn$_2$As$_3$ through two complementary mechanisms: the pre-existing in-plane ordering below $T_{\mathrm{M}}$ reduces the heat capacity contribution at $T_{\mathrm{N}}$, and the alternating mediation of inter-plane Eu-Eu coupling through CuAs planes and Zn$_2$As$_2$ layers impedes simultaneous establishment of inter-plane ordering.

Previous inelastic neutron scattering studies have established the quasi-2D nature of magnetic interactions in EuCuAs~\cite{46EuCuAsnpj}. Owing to its heterogeneous intergrowth structure, Eu$_2$CuZn$_2$As$_3$ is expected to exhibit enhanced quasi-2D magnetic behavior compared to its parent compound EuCuAs.

%and it is completely suppressed when the magnetic field exceeds 1 T. Moreover, the change in magnetic entropy is very weak when Eu$_2$CuZn$_2$As$_3$ undergoes the magnetic transition at $T_\mathrm{M}$ $\sim$ 29.3 K. Analysis of the behavior of $C_\mathrm{p}(T)$ and $S_\mathrm{mag}(T)$ suggests that the transition at 29.3 K is not a three-dimensional (3D) long-range magnetic ordering, but may be rather quasi-long-range in-plane magnetic ordering, which needs to be confirmed by further studies. 

%For the long-range magnetic ordering of Eu$^{2+}$ at $T_\mathrm{N}$ $\sim$ 18.7 K, the $C_\mathrm{p} (T)$ displays an unusual plateau-like feature below 22 K, and the value of $C_\mathrm{p}$ decreases gradually with the decreasing temperature only after cooling below 20 K. This type of transition has never been observed in layered Eu-based 122 and 111 materials, which can be attributed primarily to the intergrowth structure of Eu$_2$CuZn$_2$As$_3$. It is well established that interlayer interactions are crucial for the stability of 3D long-range magnetic ordering in layered compounds.The presence of several competing magnetic interactions in Eu$_2$CuZn$_2$As$_3$, mediated via the CuAs plane or the Zn$_2$As$_2$ layer, makes it difficult for the Eu spin to enter the ordering state. Therefore, the plateau-like behavior observed in the specific heat of Eu$_2$CuZn$_2$As$_3$ may result from weak but strongly competing interlayer magnetic interactions and the quasi-2D magnetic correlations. 

\begin{figure*}
	\includegraphics[width=\textwidth]{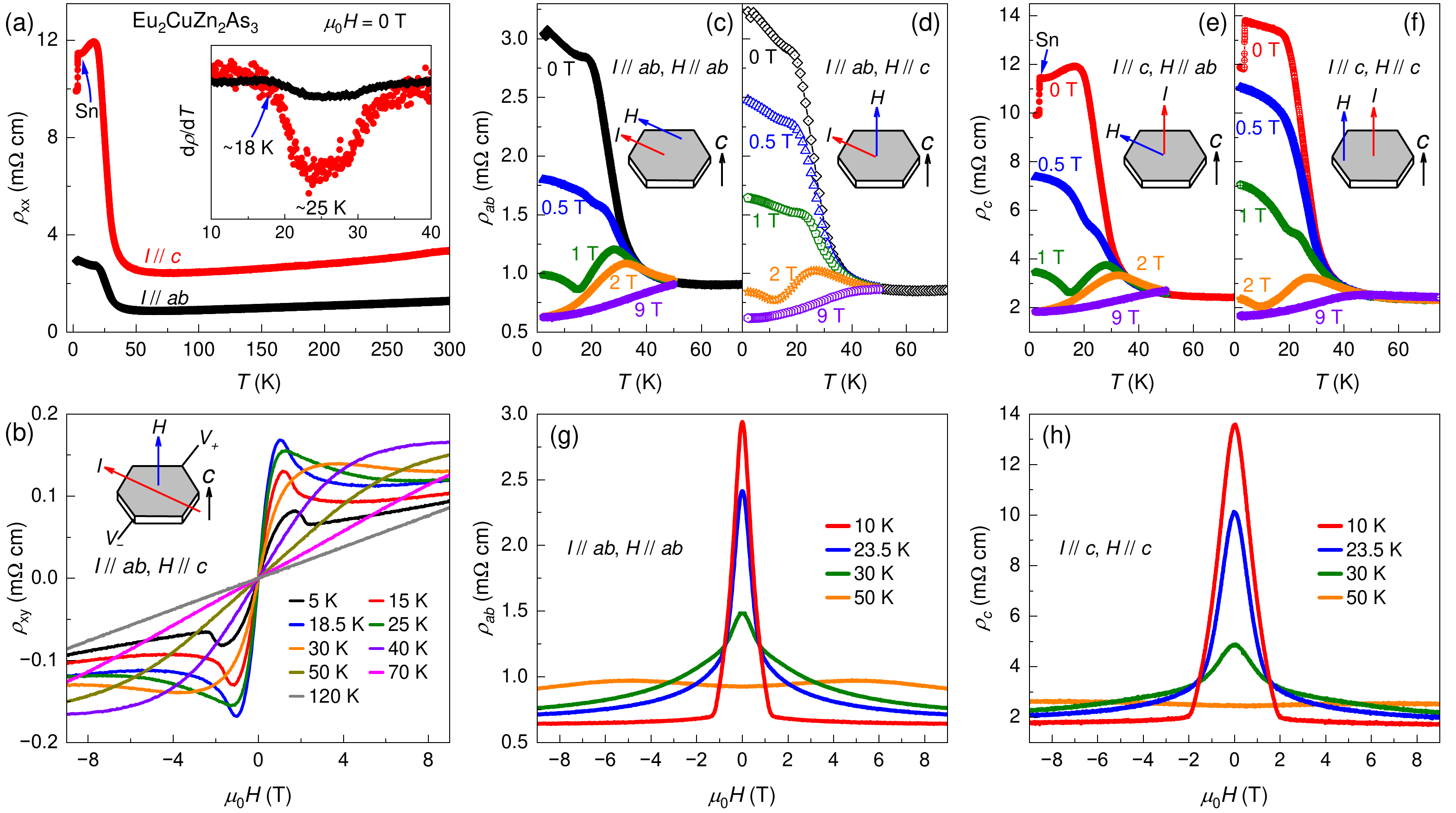}
	\caption{(a) Zero-field anisotropic resistivity $\rho_{xx}(T)$ of Eu$_2$CuZn$_2$As$_3$ with current applied along the $c$-axis ($\rho_c$, red) and $ab$-plane ($\rho_{ab}$, black). The inset shows the temperature derivative $d\rho_{xx}/dT$. (b) Field-dependent Hall resistivity $\rho_{xy}(H)$ at various temperatures. (c,d) Temperature dependence of the in-plane resistivity $\rho_{ab}(T)$ under different (c) in-plane and (d) out-of-plane magnetic fields. (e,f) Temperature dependence of the out-of-plane resistivity $\rho_c(T)$ with the magnetic field applied (e) parallel to the $ab$-plane and (f) parallel to the $c$-axis. (g,h) Magnetic field dependence of (g) $\rho_{ab}$ with $H\parallel ab$ and (h) $\rho_c$ with $H\parallel c$ measured at 10\,K, 23.5\,K, 30\,K, and 50\,K.}
	\label{F4}
\end{figure*}

\subsection{Charge transport}

Figure~\ref{F4} presents the anisotropic charge transport properties of Eu$_2$CuZn$_2$As$_3$ single crystals. As shown in Fig.~\ref{F4}(a), the zero-field resistivities $\rho_c(T)$ (current along $c$-axis) and $\rho_{ab}(T)$ (current within $ab$-plane) both exhibit metallic behavior above 50\,K with moderate magnitude variations. Below 40\,K, a rapid increase in $\rho_c$ and $\rho_{ab}$ occurs, attributed to enhanced scattering from short-range magnetic correlations preceding magnetic ordering. A distinct slope change near 18\,K, evident in the $d\rho/dT$ versus $T$ plot [inset of Fig.~\ref{F4}(a)], marks the establishment of long-range magnetic ordering, consistent with the AFM transition temperature identified from both $\chi(T)$ and $C_\mathrm{p}(T)$ measurements. The resistivity anisotropy ratio $\rho_c/\rho_{ab}$ evolves from $\sim$2.5 at room temperature to $\sim$5 below 18\,K, demonstrating enhanced anisotropy in the magnetically ordered state. The higher out-of-plane resistivity is characteristic of layered materials. Notably, unlike many Eu-based layered materials, Eu$_2$CuZn$_2$As$_3$ shows no significant resistivity drop below the ordering temperature, indicating persistent strong spin scattering~\cite{23EuZn2As2wzc,6EuCuAs_JACS}. This observation, consistent with the minimal reduction in $\chi(T)$ below $T_\mathrm{N}$, suggests the presence of substantial uncompensated spins, likely resulting from numerous AFM domain boundaries. Furthermore, the resistivity continues to increase below 18\,K, mirroring behavior reported in the parent compound EuZn$_2$As$_2$~\cite{23EuZn2As2wzc}.

Figures~\ref{F4}(c--h) present the anisotropic magnetotransport properties of Eu$_{2}$CuZn$_{2}$As$_{3}$. The in-plane resistivity $\rho_{ab}$, measured with magnetic fields applied along the $ab$-plane ($H \parallel ab$) and $c$-axis ($H \parallel c$), is shown in Figs.~\ref{F4}(c) and (d), respectively. Correspondingly, Figs.~\ref{F4}(e) and (f) display the out-of-plane resistivity $\rho_c$ under similarly oriented fields. We note that the zero-field resistivity values of both $\rho_{ab}$ [Figs.~\ref{F4}(c,d)] and $\rho_c$ [Figs.~\ref{F4}(e,f)] exhibit minor variations, attributable to measurements performed on different samples or during separate experimental runs. The results in Figs.~\ref{F4}(c--h) demonstrate a significant nMR effect and weak resistivity anisotropy, observed across different magnetic field configurations and current directions. With increasing magnetic field, magnetic scattering is rapidly suppressed below 50\,K, leading to reduced resistivity. Notably, resistivity dips appear in certain curves [e.g., the 1\,T curve in Fig.~\ref{F4}(c)], which arise from competing mechanisms involving robust spin scattering and magnetic suppression. For measurements with $H \parallel ab$ [Figs.~\ref{F4}(c,e)], these dips vanish between 1--2\,T, whereas for $H \parallel c$ [Figs.~\ref{F4}(d,f)], they persist up to 2\,T before disappearing. These critical fields are consistent with the anisotropic saturation fields in Fig.~\ref{F2}(e). Similar dip feature has been previously reported in EuCd$_{2}$As$_{2}$~\cite{4PRL_EuCu2As2,rahnCouplingMagneticOrder2018}. When comparing $\rho_{ab}(T)$ and $\rho_{c}(T)$ under identical field orientations [e.g., curves in Figs.~\ref{F4}(c) and (e)], the primary distinction lies in the resistivity magnitude, attributable to the layered crystal structure.

%The magnetic field-dependent resistivity of Eu$_2$CuZn$_2$As$_3$ is shown in Fig.~\ref{F4}(g) [$\rho_{ab}(H)$, $I \parallel H \parallel ab$] and \ref{F4}(h) [$\rho_{c}(H)$, $I \parallel H \parallel c$]. Large nMR effect are found below 50 K. For instance, the decline of $\rho_c$ is over 80\% ($\Delta\rho/\rho(0) \times 100\%$) with a 2\,T field along the $c$ axis at 10\,K in Fig.~\ref{F4}(h). The nMR effect saturates at magnetic fields exceeding 1 T and 2 T for $\rho_{ab}(H)$ and $\rho_{c}(H)$, respectively, due to the fully spin polarization above $M_\mathrm{sat}$. Additional $\rho_{xx}(H)$ curves and calculated MR are provided in the Supplemental Materials (SM). Moreover, we note a weak positive MR effect for the 50\,K curves, which is most significant for the configurations $I\parallel c$ and $H\parallel ab$ (Fig. S1(d) in the SM), resulting from enhanced scattering when the field forces the spins in the short magnetic correlations.

The magnetic field-dependent resistivity of Eu$_{2}$CuZn$_{2}$As$_{3}$ is presented in Fig.~\ref{F4}(g) [$\rho_{ab}(H)$, $I \parallel H \parallel ab$] and Fig.~\ref{F4}(h) [$\rho_{c}(H)$, $I \parallel H \parallel c$]. A pronounced nMR effect is observed below 50\,K. For example, $\rho_c$ exhibits a reduction exceeding 80\% [$\Delta\rho/\rho(0) \times 100\%$] at 10\,K under a 2\,T field applied along the $c$ axis in Fig.~\ref{F4}(h). The nMR saturation occurs at fields above 1\,T for $\rho_{ab}(H_{ab})$ and 2\,T for $\rho_{c}(H_c)$, consistent with full spin polarization above $M_\mathrm{sat}$. Additional $\rho_{xx}(H)$ curves are presented in Fig.~\ref{F7} in the Appendix. Notably, a weak positive MR effect emerges at 50~K, most prominently for the $I \parallel c$ and $H \parallel ab$ configuration [Fig.~\ref{F7}(d)]. This feature originates from enhanced scattering due to field-induced orientation of magnetically correlated spins.

%Figure \ref{F4}(b) displays the Hall resistivity $\rho_{xy}(H)$ of Eu$_2$CuZn$_2$As$_3$ with $I \parallel ab$ and $H \parallel c$ at various temperatures. The $\rho_{xy}(H)$ curves at 120 K and 70 K exhibit linear behavior with the magnetic field, indicating that the dominant ordinary Hall effect, while the positive slope suggests that the dominant carrier is the hole, consistent with carrier type reported in its parent compounds. The hole concentration of Eu$_2$CuZn$_2$As$_3$ was calculated to be $n = 6.54 \times 10^{19}$ cm$^{-3}$ with the 120\,K curve based on the single-band model. The carrier concentrations of Eu-based layered Zintl compounds are strongly affected by the crystal preparation process, and the carrier concentration could have a significant impact on the final magnetic and transport properties. The reported values of carrier concentration of EuCuAs are on the order of magnitude of $10^{20}$ cm$^{-3}$, while those of EuZn$_2$As$_2$ mostly on the order of $10^{17}$ cm$^{-3}$. Since $V_{2123}$ (the volume of the intergrowth structure) is approximately the sum of $V_{111}$ (the volume of EuCuAs) and $2\times V_{122}$ (the volume of EuZn$_2$As$_2$), it is reasonable to anticipate that the carrier concentration of Eu$_2$CuZn$_2$As$_3$ is about 1/3 of that of EuCuAs by considering the charge from the parent compounds redistributing in the intergrowth structure.

Figure~\ref{F4}(b) presents the Hall resistivity $\rho_{xy}(H)$ of Eu$_2$CuZn$_2$As$_3$ measured with $I \parallel ab$ and $H \parallel c$ at various temperatures. The $\rho_{xy}(H)$ curves at 120\,K and 70\,K display linear field dependence, indicating dominant ordinary Hall effect. The positive slope confirms hole-type carriers as majority charge carriers, consistent with reports for its parent compounds. Based on the single-band model, the hole concentration at 120\,K is calculated to be $n = 6.54 \times 10^{19}$\,cm$^{-3}$. Carrier concentrations in Eu-based layered Zintl phases are highly sensitive to crystal growth conditions and substantially influence resultant magnetic and transport properties~\cite{chenRecentAdvancesUnderstanding2024}. The reported carrier concentrations of EuCuAs are typically on the order of $10^{20}$\,cm$^{-3}$, while those of EuZn$_2$As$_2$ are mostly on the order of $10^{17}$\,cm$^{-3}$. Considering the approximate volume relationship $V_{\mathrm{2123}} \approx V_{\mathrm{111}} + 2V_{\mathrm{122}}$%(where subscripts denote Eu$_2$CuZn$_2$As$_3$, EuCuAs, and EuZn$_2$As$_2$ structures respectively)
, and assuming charge redistribution within the intergrowth framework, the carrier concentration in Eu$_2$CuZn$_2$As$_3$ is expected to approach $\sim$1/3 of that in EuCuAs, which is consistent with our fitted hole density.

With decreasing temperature, the $\rho_{xy}(H)$ curve at 50\,K develops a discernible curvature, indicating a contribution from the anomalous Hall effect (AHE) due to short-range magnetic correlations. Below 30\,K, an additional peak emerges in the $\rho_{xy}(H)$ curves, intensifying progressively and reaching a maximum at 18.5\,K. This peak feature exhibits nonlinear dependence on both magnetic field and magnetization, characteristic of the NLAHE frequently observed in Eu-based materials such as EuCuAs and EuZn$_2$As$_2$~\cite{6EuCuAs_JACS,46EuCuAsnpj,24EuZn2As2LuoShuaishuai,47EuZn2As2THE}. The NLAHE is understood to originate from the presence of topological magnetic or electronic textures, both of which represent plausible mechanisms for Eu$_2$CuZn$_2$As$_3$. In real space, non-zero Berry curvature may arise from finite scalar spin chirality within domain walls, as demonstrated in EuZn$_2$As$_2$, EuZn$_2$Sb$_2$, and EuAl$_2$Si$_2$~\cite{47EuZn2As2THE,17EuZn2Sb2,xiaGiantDomainWall2024}. In momentum space, it could stem from a topologically nontrivial band structure. Previous studies have established that EuCuAs in both helical magnetic and FM polarized states can host Weyl nodes~\cite{46EuCuAsnpj}, while EuZn$_2$As$_2$ can be driven into a Weyl metal state through applied field or pressure~\cite{24EuZn2As2LuoShuaishuai}. The structural homology with its parent compounds suggests a high probability of nontrivial band structure in Eu$_2$CuZn$_2$As$_3$. Indeed, our calculations indicate the presence of potentially nontrivial band crossings near the Fermi level [Figs.~\ref{F6}(a-c)], which will be analyzed later.

\subsection{Magnetic energies}

\begin{figure}
	\includegraphics[width=0.48\textwidth]{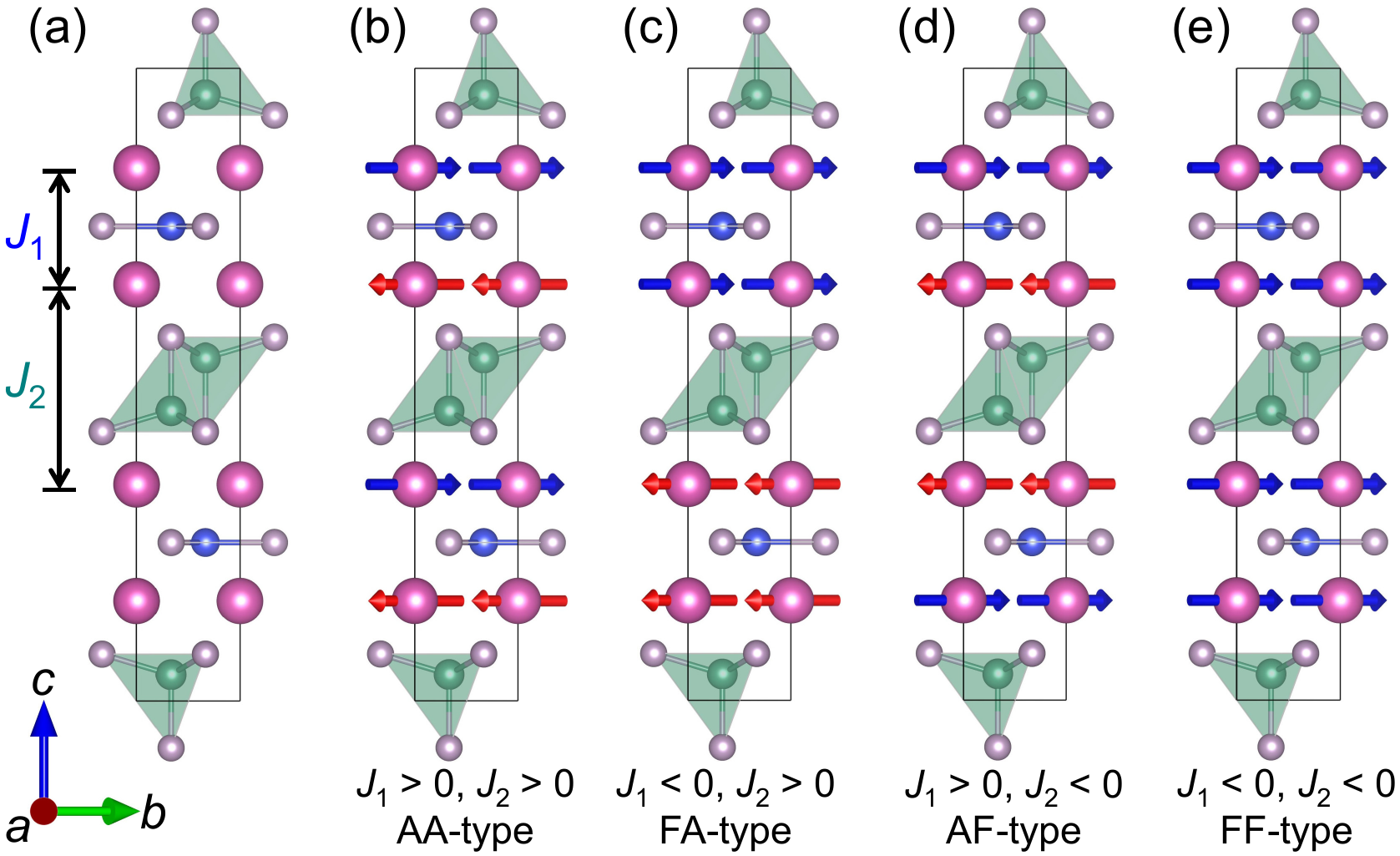}
	\caption{(a) Key interplane Eu-Eu couplings mediated by CuAs planes ($J_1$) and Zn$_2$As$_2$ layers ($J_2$). Positive coupling coefficients denote AFM (A) interactions, while negative values indicate FM (F) coupling. (b-e) Proposed magnetic configurations for Eu$_2$CuZn$_2$As$_3$: (b) AA-type ($J_1>0$, $J_2>0$), (c) FA-type ($J_1<0$, $J_2>0$), (d) AF-type ($J_1>0$, $J_2<0$), (e) FF-type ($J_1<0$, $J_2<0$).}	
	\label{F5}
\end{figure}

	%%%%%%%%%%%%%%%%%%% TABLE2 %%%%%%%%%%%%%%%%%%%%%
\begin{table}
	\caption{Calculated magnetic energies (without spin-orbit coupling, SOC) per formula unit (meV/f.u.) for Eu$_2$CuZn$_2$As$_3$ corresponding to the spin configurations proposed in Fig.~\ref{F5}, along with the optimized lattice constants. Corresponding Eu magnetic moments remain approximately constant at 6.95 $\mu_\mathrm{B}$/Eu across all magnetic configurations. The magnetic energy of the AA-type structure is normalized to zero, with the energy values of other configurations expressed as relative values to it.
	}
	
	\begin{ruledtabular}\label{Tab-2}
		\begin{tabular}{lcccc}	
			 & AA-type & FA-type & AF-type & FF-type  \\
			\hline
			$a$ (\AA) & 4.190 & 4.189 & 4.190 & 4.190 \\
			$c$ (\AA) & 22.299 & 22.301 & 22.310 & 22.304 \\
			$E$ (meV/f.u.) & 0 & $-1.65$ & 1.37 & 0.50 
		\end{tabular}
	\end{ruledtabular}
\end{table}

%能量差是小的
%印证了铁磁关联的存在in the 2123

%We calculated the magnetic energies of several possible collinear spin configurations for Eu$_2$CuZn$_2$As$_3$ to explore its magnetic ground state. The selected magnetic structures, as shown in Fig.~\ref{F5}, are distinguished by the interplane Eu-Eu coupling signs. Both AFM and FM couplings are considered for the neighboring Eu planes across CuAs planes or Zn$_2$As$_2$ layers. Here we only considered the collinear magnetic structures, given the challenges of calculating the noncollinear AFM order in the intergrowth structure, although the a helical spin order was reported for the parent material EuCuAs. Consequently, four types of the spin structures are considered, with the nomenclature of these structures explained in the caption of Fig.~\ref{F5}, while the corresponding energy values are listed in Table~\ref{Tab-2}. Our calculations that the FA-type AFM order, with FM coupling for the Eu planes within the EuCuAs units and AFM coupling for the EuZn$_2$As$_2$ units, shows the lowest energy. This spin configuration is identical to the magnetic structure of Eu$_2$CuZn$_2$P$_3$, which is determined by the single crystal neutron diffraction. Hence, we conjecture that the FA-type AFM order is the likely ground state of Eu$_2$CuZn$_2$As$_3$. However, the possibility that Eu$_2$CuZn$_2$As$_3$ possesses a noncollinear AFM order cannot be simply ruled out, and further studies using neutron diffraction and resonant elastic x-ray scattering are encouraged to examine this possibility.

To determine the magnetic ground state of Eu$_2$CuZn$_2$As$_3$, we calculated the energies of several possible collinear spin configurations. The magnetic structures under consideration (Fig.~\ref{F5}) are characterized by their interplane Eu-Eu coupling signs, including both AFM and FM interactions across CuAs planes and Zn$_2$As$_2$ layers. Although the parent compound EuCuAs exhibits a helical spin order~\cite{6EuCuAs_JACS,46EuCuAsnpj}, our investigation was restricted to collinear configurations due to computational challenges in modeling noncollinear AFM ordering within intergrowth structures.

Four distinct spin structure types were evaluated (see Fig.~\ref{F5} caption for nomenclature), with corresponding energy values listed in Table~\ref{Tab-2}. Our calculations demonstrate that the FA-type AFM order, featuring FM coupling between Eu planes within EuCuAs units and AFM coupling within EuZn$_2$As$_2$ units, exhibits the lowest energy. This configuration matches the magnetic structure of Eu$_2$CuZn$_2$P$_3$ determined by single-crystal neutron diffraction~\cite{44Eu2CuZn2P3_PRM}. We therefore propose the FA-type AFM order as the probable ground state for Eu$_2$CuZn$_2$As$_3$. Nevertheless, the potential existence of a noncollinear AFM ordering cannot be definitively excluded based on our calculations alone. Additional experimental verification through neutron diffraction and resonant elastic x-ray scattering measurements is warranted.

%Since an easy-plane spin arrangement has been manifested by the magnetic measurements, we only consider the cases that spins lie in the ab

%The magnet

%In order to gain a deeper understanding of the magnetic properties of Eu$_2$CuZn$_2$As$_3$, we calculated various possible spin configurations for the identification of its magnetic ground state. In consideration of the collinear magnetic structure in EuZn$_2$As$_2$, four collinear spin configurations for Eu$_2$CuZn$_2$As$_3$ are proposed, as shown in Fig. \ref{F5}(a). Although EuCuAs with a helical spin order, given the relative difficulty of calculating this configuration, the noncollinear spin structure is not considered in our alternative. Studies using neutron diffraction and resonant elastic x-ray scattering is encouraged to further examine the possibility of noncollinear spin order in Eu$_2$CuZn$_2$As$_3$. Our calculations suggest that the AFM-2 spin structure with the lowest energy, this spin configuration was recently identified as the magnetic ground state of, a sister compound, Eu$_2$CuZn$_2$P$_3$. Consequently, the proposed AFM-2 configuration is plausible for Eu$_2$CuZn$_2$As$_3$, and its corresponding electronic band structure is shown in Fig. \ref{F5}(b).

\subsection{Electronic structure}

%因为具体的磁结构还未确定，所以只能用猜测的来计算
%费米面DOS低，与低载流子浓度相符
%金属性
%多带性
%能带交叉
%semimetal
%reveals a low density of electronic states near the Fermi energy EFThis is consistent with the semimetallic nature of the compound as suggested by transport measurements PHYSICAL REVIEW B 98, 064419 (2018)
%The rich interplay between the magnetism and band topology in materials ensures a variety of topological states.
%轨道杂化
\begin{figure*}
	\includegraphics[width=\textwidth]{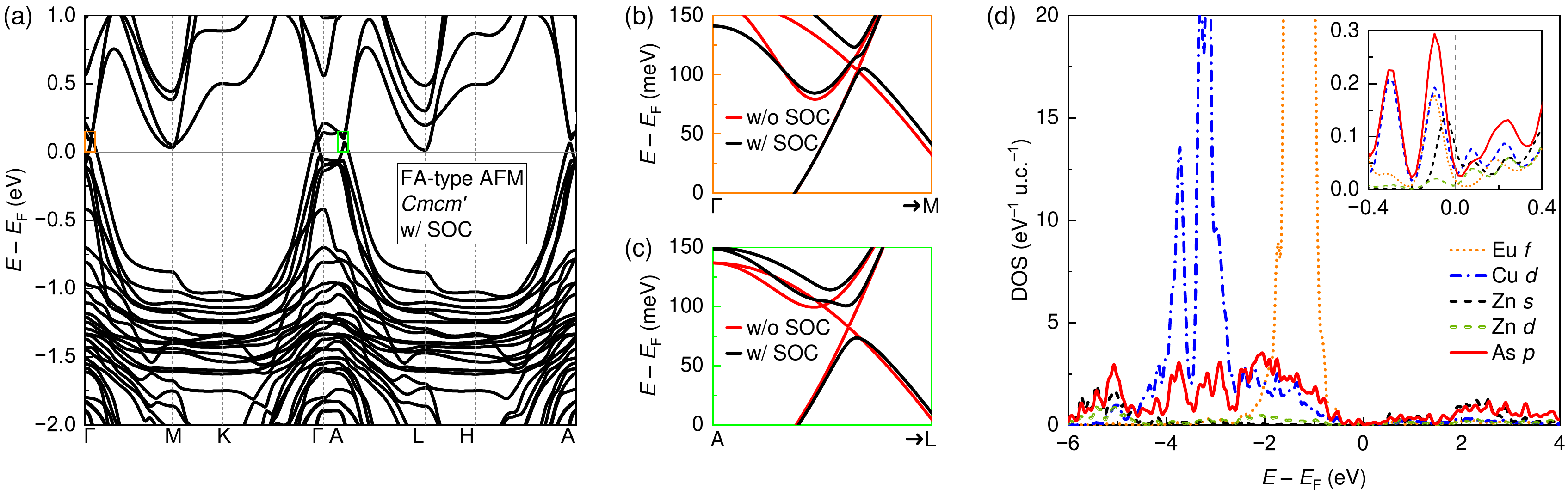}
	\caption{(a) Calculated band structure of FA-type AFM Eu$_2$CuZn$_2$As$_3$ with SOC included (moment along $a$ axis). (b,c) Band dispersions along high-symmetry paths (b) $\Gamma$-M and (c) A-L. Black and red curves denote calculations with and without SOC, respectively. (d) DOS for Eu$_2$CuZn$_2$As$_3$. The inset magnifies specific orbital-projected DOS near the Fermi level.}
	\label{F6}
\end{figure*}

We present the calculated band structure and density of states (DOS) of Eu$_2$CuZn$_2$As$_3$ in the FA-type AFM configuration, corresponding to magnetic space group $Cmcm^\prime$, as shown in Fig.~\ref{F6}. The calculations reveal a low electronic DOS near the Fermi level ($E_\mathrm{F}$) [Fig.~\ref{F6}(d)], consistent with the low carrier density inferred from Hall resistivity measurements. The Eu 4$f$ states are predominantly localized between 1.5 and 2 eV below $E_\mathrm{F}$, indicating strong electronic localization.

In contrast to Eu$_2$CuZn$_2$P$_3$, whose band calculations reveal a narrow-gap semiconductor behavior under identical spin configuration~\cite{44Eu2CuZn2P3_PRM}, Eu$_2$CuZn$_2$As$_3$ exhibits band inversion between Zn $s$- and As $p$-derived orbitals near the $\Gamma$ and A high-symmetry points. Comparative analysis of band dispersions with and without SOC is presented along $\Gamma$-M and A-L paths in Figs.~\ref{F6}(b) and \ref{F6}(c), respectively. In the absence of SOC, symmetry-protected band crossings persist due to the preservation of both inversion symmetry ($\mathcal{P}$) and effective time-reversal symmetry $\mathcal{T}^\prime \equiv \mathcal{T}\otimes\tau$ (where $\tau$ is a nonsymmorphic translation), characteristic of an AFM Dirac semimetal phase.

The inclusion of SOC with moments oriented in the $ab$-plane breaks the in-plane sixfold rotational symmetry, opening band gaps of tens of meV. This symmetry-reducing-induced gapping mechanism mirrors observations in related compounds EuCd$_2$As$_2$ and EuCuAs~\cite{rahnCouplingMagneticOrder2018,huaDiracSemimetalTypeIV2018,32WangLin-Lin_EuCd2As2,6EuCuAs_JACS}. While a complete mapping of topological states across all spin configurations exceeds the scope of this study, the established magneto-topological coupling in Eu$M_2X_2$ and Eu$TX$ systems suggests that Eu$_2$CuZn$_2$As$_3$ should host multiple spin-dependent topological phases, warranting detailed future investigation~\cite{46EuCuAsnpj,jinMultipleMagnetismcontrolledTopological2021,lahaTopologicalHallEffect2021,ramMagnetotransportElectronicStructure2024,malickElectronicStructurePhysical2022,yuanMagnetizationDependentAnisotropic2024}.

\section{CONCLUSIONS}
%In summary, single crystals of Eu$_2$CuZn$_2$As$_3$ were successfully prepared via Sn flux, and we systematically investigated its magnetic and transport properties through a combination of experimental characterizations and theoretical calculations. As an intergrowth phase of EuCuAs and EuZn$_2$As$_2$, Eu$_2$CuZn$_2$As$_3$ exhibits not only physical properties comparable to those of parent compounds, but also emerges unique features. Eu$_2$CuZn$_2$As$_3$ exhibit two stepwise AFM transitions at $T_\mathrm{M}$ = 29.3\,K and $T_\mathrm{N}$ = 19 K, as well as a spin-reorientation transition at $T_\mathrm{SR}$ = 16.3 K. Interestingly, magnetic transitions at 18.7 K and 16.3\,K are very close to that of the EuZn$_2$As$_2$ (19.6 K) and EuCuAs (16 K). This similarity suggests that  the magnetic transitions at $T_\mathrm{N}$ and $T_\mathrm{SR}$ of Eu$_2$CuZn$_2$As$_3$ arise from magnetic interactions mediated via Zn$_2$As$_2$ layers and CuAs planes, respectively. A similar scenario was recently observed in the sibling compound Eu$_2$CuMn$_2$P$_3$. However, Eu$_2$CuZn$_2$P$_3$, the intergrowth compound of EuCuP and EuZn$_2$P$_2$, exhibits only one magnetic transition that is distinct from its parent materials. This discrepancy between the above intergrowth phases is intriguing and may be due to different the strength of various competing magnetic correlations or spin-orbit coupling effect in the compounds.

In summary, single crystals of Eu$_2$CuZn$_2$As$_3$ were successfully synthesized using the Sn flux method. Through comprehensive experimental characterization and theoretical calculations, we systematically investigated the compound's magnetic and transport properties, as well as the band structure. As an intergrowth phase of EuCuAs and EuZn$_2$As$_2$, Eu$_2$CuZn$_2$As$_3$ exhibits $c$-axis heterogeneity that modulates magnetic correlation strengths, resulting in stepwise AFM transitions. The compound undergoes two conspicuous AFM transitions at $T_\mathrm{M} = 29.3$\,K and $T_\mathrm{N} = 19$\,K, followed by a spin-reorientation transition at $T_\mathrm{SR} = 16.3$\,K. The stepwise transitions are manifested in the heat capacity as unconventional plateau-like anomalies. Notably, the transitions at 19\,K and 16.3\,K closely correspond to those in EuZn$_2$As$_2$ (19.6\,K) and EuCuAs (16\,K), respectively. This similarity strongly suggests that the $T_\mathrm{N}$ and $T_\mathrm{SR}$ transitions in Eu$_2$CuZn$_2$As$_3$ originate from magnetic interactions mediated by the Zn$_2$As$_2$ layers and CuAs planes, respectively. Such behavior has been recently observed in the analogous compound Eu$_2$CuMn$_2$P$_3$. In contrast, the isostructural phosphide Eu$_2$CuZn$_2$P$_3$, an intergrowth of EuCuP and EuZn$_2$P$_2$, exhibits a primary magnetic transition at higher temperature than either parent compound, along with a weak anomaly resembling spin reorientation. This intriguing discrepancy between intergrowth phases may stem from differences in the relative strengths of competing magnetic correlations or SOC effects.

Transport measurements on Eu$_2$CuZn$_2$As$_3$ reveal semimetallic behavior at high temperatures and a significant resistivity increase below 40\,K. Unlike its parent compounds, Eu$_2$CuZn$_2$As$_3$ does not exhibit the characteristic resistivity drop below the AFM ordering temperature, which we attribute to the presence of substantial uncompensated spins persisting in the ordered state. Under applied magnetic fields, large nMR and a significant NLAHE emerge, similar to observations in EuCuAs and EuZn$_2$As$_2$. To investigate the spin structure of Eu$_2$CuZn$_2$As$_3$, we calculated the magnetic energies for several collinear configurations. Our results indicate that the most stable ground state features AFM interplane Eu-Eu couplings within the EuZn$_2$As$_2$ units and FM couplings within the EuCuAs units (FA-type configuration). This spin structure is consistent with previous reports for Eu$_2$CuZn$_2$P$_3$. However, given the helical spin arrangement observed in EuCuAs, future studies using neutron diffraction or other techniques should investigate whether Eu$_2$CuZn$_2$As$_3$ also exhibits noncollinear spin ordering. Furthermore, our band structure calculations for the FA-type spin configuration suggest that the FA-type spin configuration may host topologically nontrivial band crossings, consistent with the observed large NLAHE. Further investigations exploring the topological states in Eu$_2$CuZn$_2$As$_3$ and their coupling to magnetic order are of significant interest.

Our work demonstrates a structural design strategy for synthesizing topologically nontrivial intergrowth materials by hybridizing known layered topological compounds. This approach shows great potential for discovering additional intergrowth materials with novel properties. We anticipate that more 2123-type intergrowth compounds will be identified through combinations of CaAl$_{2}$Si$_{2}$-type and SrPtSb-type materials in future studies.

%\alert{The low-filed region $M_c(H)$ curves is shown in SI, all of which exhibit linear behavior and the data from 5 to 20 K almost overlap.}

%\alert{Considering the unusual magnetic and transport properties manifested in Eu$_2$CuZn$_2$As$_3$, further studies on its magnetic and electronic structure using neutron diffraction and angle-resolved photoemission spectroscopy are deserving in the future.}

\begin{acknowledgments}
This work was supported by the National Natural Science Foundation of China (Grants No. 12204094, No. 12325401, and No. 123B2053), the Natural Science Foundation of Jiangsu Province (Grant No. BK20220796), the Start-up Research Fund of Southeast University (Grant No. RF1028623289), the SEU Innovation Capability Enhancement Plan for Doctoral Students (Grant No.CXJH\_SEU 25136), and the Big Data Computing Center of Southeast University.
\end{acknowledgments}

\section*{DATA AVAILABILITY}
The crystallographic data for Eu$_2$CuZn$_2$As$_3$ are openly available~\cite{CCDC}. All other data supporting the findings of this study are not publicly available upon publication because it is not technically feasible and/or the cost of preparing, depositing, and hosting the data would be prohibitive within the terms of this research project. The data are available from the authors upon reasonable request.

\appendix*
\section{Additional $\rho_{xx}(H)$ curves}

\begin{figure*}
	\includegraphics[width=0.75\textwidth]{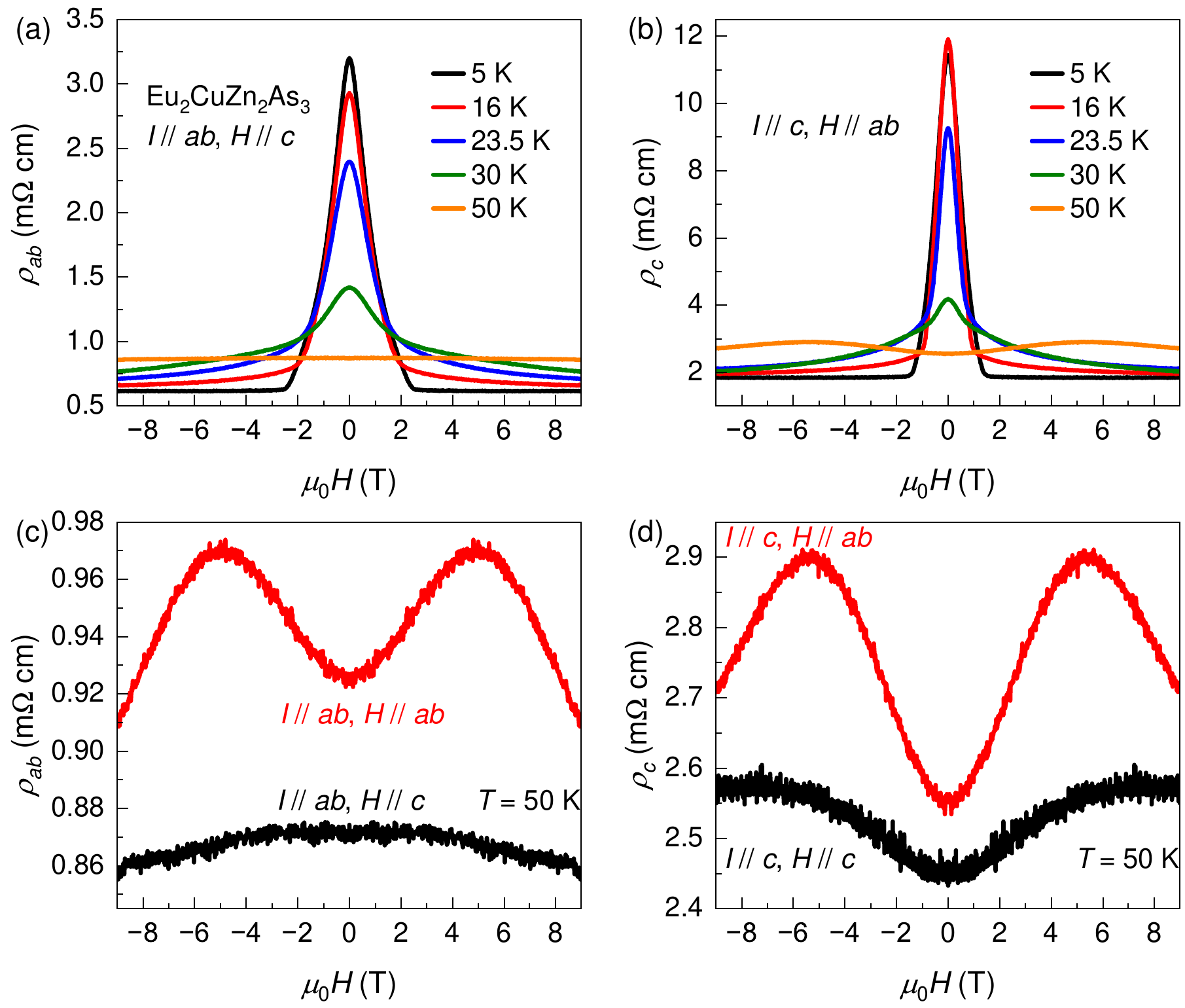}
	\caption{(a,b) Field-dependent resistivity curves for (a) $\rho_{ab}$ with $H \parallel c$ and (b) $\rho_{c}$ with $H \parallel ab$ at selected temperatures (5, 16, 23.5, 30, and 50\,K). (c,d) Comparative analysis at 50 K: (c) $\rho_{\mathrm{ab}}$ under  $H \parallel ab$ versus $H \parallel c$; (d) $\rho_{c}$ under $H \parallel ab$ versus $H \parallel c$.}

	\label{F7}
\end{figure*}

Complementary magnetoresistivity data are provided in Fig.~\ref{F7}.  

\bibliography{Eu2CuZn2As3refs}
\end{document}